\documentclass[useAMS]{mn2e}
\usepackage{graphicx}
\usepackage{longtable}

\title[Star formation in H{\sc i} tails]{Star formation in H{\sc i} tails: HCG~92, HCG~100 and 6 interacting systems\thanks{Based on observations obtained at the Gemini Observatory, which is operated by the Association of Universities for Research in Astronomy, Inc., under a cooperative agreement with the NSF on behalf of the Gemini partnership: the National Science Foundation (United States), the Science and Technology Facilities Council (United Kingdom), the National Research Council (Canada), CONICYT (Chile), the Australian Research Council (Australia), Minist\'erio da Ci\^ encia e Tecnologia (Brazil), and Ministerio de Ciencia, Tecnologia e Innovacion Productiva (Argentina -- Observing run ID: GN-2003A-Q-53 and GN-2007B-Q-87.}}

\author[de Mello et al.]
{
\parbox[t]{\textwidth} {D. F. de Mello$^{1,2}$\thanks{E-mail: demello@cua.edu}, F. Urrutia-Viscarra$^{3}$, C. Mendes de Oliveira$^{3}$, S. Torres-Flores$^{4}$, E. R. Carrasco$^{5}$, E. Cypriano $^{3}$}
\vspace*{6pt}\\
$^{1}$The Catholic University of America, Physics Department, Washington, DC 20064, USA\\
$^{2}$Observational Cosmology Laboratory, Code 665, Goddard Space Flight Center, Greenbelt, MD 20771, USA\\
$^{3}$Departamento de Astronomia, Instituto de Astronomia, Geof\'isica e Ci\^encias Atmosf\'ericas da USP,\\ Rua do Mat\~ao 1226, Cidade Universit\'aria, 05508-090, S\~ao Paulo, Brazil\\
$^{4}$Departamento de F\'isica, Universidad de La Serena, Av. Cisternas 1200 Norte, La Serena, Chile \\
$^{5}$Gemini Observatory/AURA, Southern Operations Center, Casilla 603, La Serena, Chile
}

\begin{document}

\date{}

\pagerange{\pageref{firstpage}--\pageref{lastpage}} \pubyear{2010}

\maketitle

\label{firstpage}

\begin{abstract}
We present new Gemini spectra of 14 new objects found within the H{\sc i} tails of  Hickson Compact Groups 92 and 100. Nine of them are GALEX Far-UV (FUV) and Near-UV (NUV) sources. The spectra confirm that these objects are members of the compact groups and have metallicities close to solar, with an average value of 12+log(O/H)$\sim$8.5. They have average FUV luminosities  7 $\times$ 10$^{40}$ erg s$^{-1}$, very young ages ($<$ 100 Myr) and two of them resemble tidal dwarf galaxies (TDGs) candidates. We suggest that they were created within gas clouds that were ejected during galaxy-galaxy interactions into the intergalactic medium, which would explain the high metallicities of the objects, inherited from the parent galaxies from which the gas originated. We conduct a search for similar objects in 6 interacting systems with extended H{\sc i} tails, NGC~2623, NGC~3079, NGC~3359, NGC~3627, NGC~3718, NGC~4656. We found 35 UV sources with ages $<$ 100 Myr, however most of them are on average less luminous/massive than the UV sources found around HCG~92 and 100. We speculate that this might be an environmental effect and that compact groups of galaxies are more favorable to TDG formation than other interacting systems.


\end{abstract}


\section{Introduction}

Interacting galaxies are ideal laboratories to probe galaxy evolution since tidal interaction
is an important mechanism in shaping galaxies properties as we measure today. 
The H{\sc i} gas, which is both the reservoir for star formation and an
excellent tracer of the large-scale galaxy dynamics, is affected by tidal interaction and is
often found in tails outside interacting galaxies.  
One of the key questions regarding the encounters of disk galaxies is the fate of the stripped H{\sc i} gas. Do these H{\sc i} intergalactic clouds form new stellar systems and/or dwarf galaxies known as tidal dwarf galaxies (TDGs)? 
And if they do, is there any difference in the types of objects that could be formed based on the type of environment where they are located? We have embarked in a series of papers trying to answer theses questions. In Torres-Flores et al. (2009), de Mello et al. (2008a) and Mendes de Oliveira et al. (2004, 2006) we showed a few Hickson Compact Groups (HCGs) 
 contain TDGs and intragroup star-forming regions. Other authors have also found {\bf TDGs} and many young globular cluster candidates  in compact groups (e.g., Iglesias-P\'aramo $\&$ V\'ilchez 2001 and Gallagher et al. 2001). Other cases of intergalactic star-forming regions have also been reported outside interacting galaxies (e.g. Ryan-Weber et al. 2004, Mullan et al. 2011, Oosterloo et al. 2004, Werk et al. 2011), including young ($<$10 Myr) small stellar clusters in the H{\sc i} bridge between M81 and M82 (de Mello et al. 2008b) and outside the merger remnant NGC2782 (Torres-Flores et al. 2012).

The importance of these newly formed objects as products of collisions is still debatable. They may be responsible for enriching the intragroup medium with metals which may have broad implications for galaxy chemodynamical evolution (Werk et al. 2011). They could grow to become independent objects as dwarf galaxies, or live as stellar clusters in the distant halos of their hosts. In addition, one cannot exclude the possibility that they will dissolve and not remain gravitationally bound, yielding only very sparse star streams, or fall back onto the progenitor (Bournaud \& Duc 2006). 


UV images of tidal tails, obtained with the Galaxy Evolution Explorer (\textit{GALEX}) satellite, showed UV-bright regions coincident with H{\sc i} density enhancements (Hibbard et al. 2005, Neff et al. 2005). More recently, Thilker et al. (2009) reported the discovery of massive star formation in the Leo primordial H{\sc i} ring which is having one of its first bursts of star formation. Therefore, UV and H{\sc i} data together provide a powerful technique for identifying and studying star forming regions in the vicinity of interacting galaxies. In de Mello et al. (2008a) we presented a sample of 16 star forming region candidates in the intergalactic medium surrounding HCG~100. Here we present the optical data obtained with Gemini for HCG~100 and also for another compact group, HCG~92. We also present the UV data of six interacting galaxies with H{\sc i}  tails where we discovered 35 stellar cluster candidates. 
 
This paper is organized as follows: \S 2 presents the data and results for HCG~92 and 100; \S 3 presents the comparison sample and the discussion; \S 4 presents the Summary, and the Appendix describes the comparison sample in more detail. Throughout the paper we assumed $\Omega_M$ = 0.3, $\Omega_{\Lambda}$ = 0.7, and H$_{\rm 0}$ = 100 h km s$^{-1}$ Mpc$^{-1}$, with h=0.71.

\section{HCG~92 and HCG~100}

The targets analyzed here are newly identified members of two Hickson compact groups of galaxies, HCG~92 and HCG~100 (Hickson 1982) located within their H{\sc i} tails. HCG~92, known as Stephan's quintet (e.g. Moles et al. 1997, Sulentic et al. 2001, Gallagher et al. 2001), is formed by three late-type galaxies (NGC~7318a, 7318b, 7319), one early-type (NGC~7317) at 80~Mpc and one object, NGC~7320, which is a foreground object. Mendes de Oliveira et al. (2004) presented the discovery of four intergalactic H{\sc ii} regions in the H{\sc i} tail of HCG~92 located more than 25 kpc from the nearest group galaxy. 

HCG~100 is formed by a bright central Sb galaxy (HCG~100a), an irregular galaxy with an optical tidal tail (HCG~100b), a late-type barred spiral (HCG~100c) and a late-type edge-on spiral (HCG~100d). It is the last group of Hickson's catalog (1982) and is at 76.3~Mpc (v$_{R}$=5,336 km/s). 
de Mello et al. (2008a) presented GALEX FUV and NUV images of this group
and identified 15 FUV sources located in the vicinity of the intergalactic H{\sc i} clouds of
the compact group which extends to over 130 kpc away from the main galaxies.

\subsection{Spectroscopy with Gemini/GMOS}

We have obtained new spectra of dozens of UV sources identified within the H{\sc i} tails of HCG~92 and HCG~100 and derived their radial velocities. We also determined metallicities for those which turned out to be at the same redshift of the groups. Observations were performed with the Gemini Multiobject Spectrograph (GMOS) at Gemini North in June 2003 (HCG~92) and in October and November 2007 (HCG~100). We centered GMOS slit on members of the groups and on sources which were identified in Mendes de Oliveira et al. (2004) and de Mello et al. (2008a). Other objects in the field were also observed when there was space left in the multislit mask. The spectra were acquired using the B600 and R400 gratings. 

Exposure times for HCG~92 were 3$\times$1500 sec for the B600 granting and 3$\times$1000 sec for the R400 grating, covering from 3700 to 8000 \AA. For HCG~100 data, the total exposure times were 3$\times$600 sec and 3$\times$1200 sec for the B600 and R400 gratings, respectively, and the final spectra covered a wavelength interval of 3700 to 7000 \AA. Position angles were 20 and 300 degrees, from the usual orientation of GMOS, values for the airmass were 1.22 (R400) and 1.08 (B600) for HCG~92 and 1.03 for HCG~100, respectively. The seeing of 1 arcsec matched well the slit size of 1 arcsec in both cases. 

All spectra were biased, trimmed, flat fielded, and wavelength calibrated with the Gemini IRAF package version 1.8 inside IRAF\footnote{IRAF is distributed by the National Optical Astronomy Observatories, which are operated by the Association of Universities of Research in Astronomy, Inc., under cooperative agreement with the National Science Foundation.}. The final spectra have typical resolutions of 3.2 and 7.0 \AA{} for B600 and R400 gratings, respectively. The spectra of the regions in HCG~92 were not flux calibrated given that there were no standard calibrators observed around the time the data were taken. While the regions of HCG~100 had their flux calibrated using the spectrum of the stars BD+284211 (R400) and Hiltner 600 (B~600), observed in December 11, 2007. For reddening correction we used the intrinsic H$\alpha$/H$\beta$ ratio, with an intrinsic value taken by Osterbrock (2006) for an effective temperature of 10000K and N$_e$=10$^2$. 

We found 12 and 2 sources at the same redshift of HCG~92 and HCG~100, respectively. Four of the 12 sources had already been confirmed as members of HCG~92 by Mendes de Oliveira et al. (2004). 


\begin{figure*}
\centering
\includegraphics[scale=0.6]{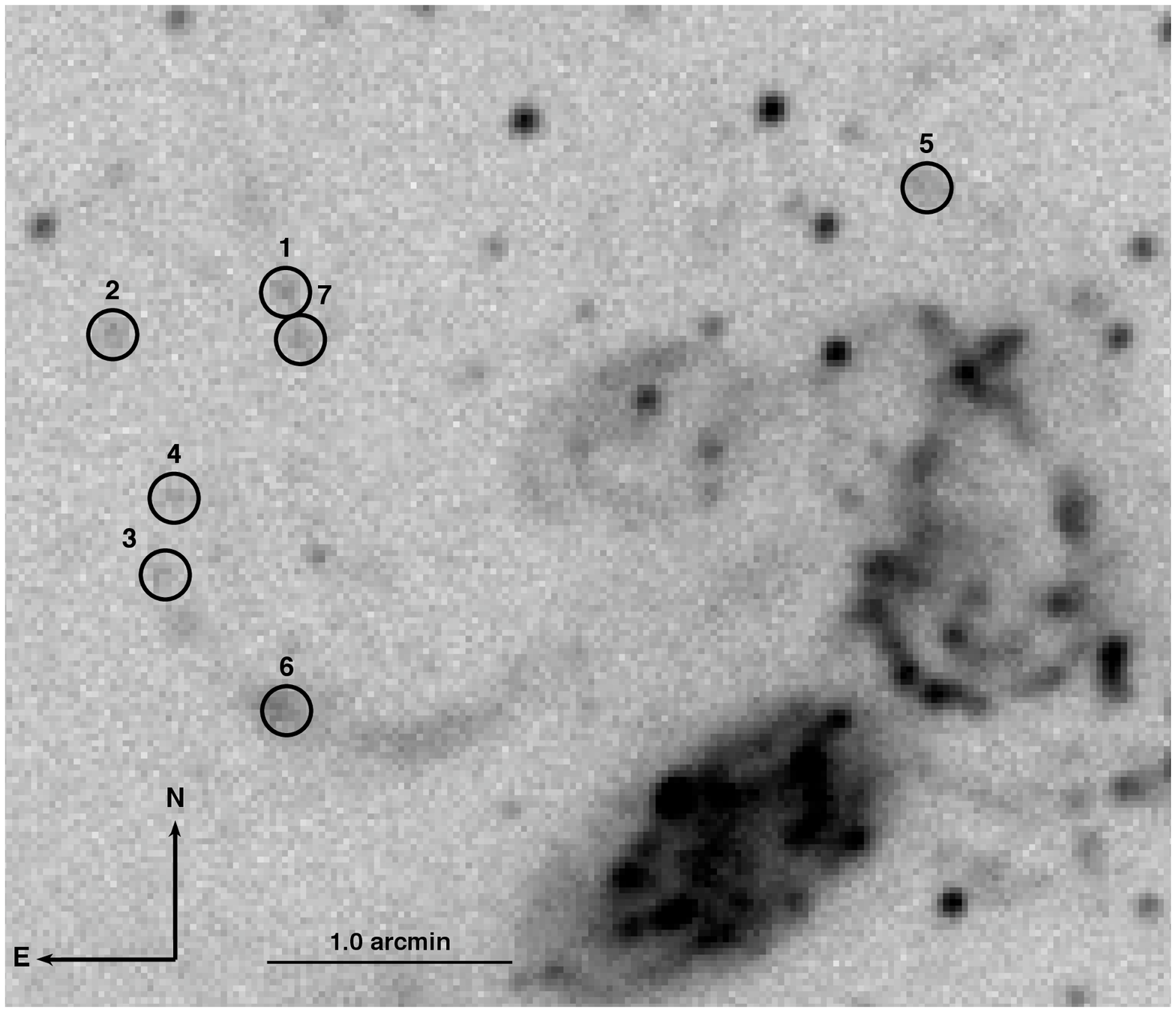}
\caption{NUV image of HCG~92. Seven UV sources detected in the GALEX images are marked. North is up and East is to the left. Bar length is 1$'$.}\label{fig1}
\end{figure*}

\subsection{GALEX data}

We obtained GALEX FUV and NUV background subtracted images from the Multimission Archive at the Space Telescope Science Institute (MAST) and followed the method by de Mello et. al (2008a) to select UV sources  within the H{\sc i} tail or in the outskirts of the H{\sc i} map.
FUV and NUV fluxes were calculated using Morrissey et al. (2005) m$_{\lambda}$=-2.5log[F$_{\lambda}$/a$_{\lambda}$] + b$_{\lambda}$. Fluxes were multiplied by the effective filter bandpass (FUV= 1528 $\pm$ 269 \AA\ and NUV = 2271 $\pm$ 616 \AA) to give units of erg s$^{-1}$ cm$^{-2}$. 

The GALEX fields of view are 1$^{\rm o}$.28 and 1$^{\rm o}$.24 in FUV and NUV, respectively, and the pixel scale is 1.5 arcsec pixel$^{-1}$. The images have a resolution (FWHM) of 4.2$\arcsec$and 5.3$\arcsec$ in FUV and NUV, respectively. Despite the broad FWHM, GALEX is able to detect faint UV sources. The medium imaging survey, for instance, reaches m=24 and 24.5 in FUV and NUV with typical exposures of 1,500s (Bianchi et al. 2007). GALEX images have also been used extensively to search for very low surface brightness objects (e. g. Thilker et al. 2007) such as the ones we are interested in detecting. We chose the parameters to detect the UV with Source Extractor and perform photometry  (version 2.4.3, Bertin $\&$ Arnouts 1996, hereafter SE) following the prescription of de Mello et al. (2008a, 2008b) which was fine tuned for detecting low surface brightness objects and clumpy systems. We matched both catalogs, FUV and NUV, within 3$\arcsec$-4$\arcsec$ radius. The SE's UV magnitudes (Mag$\_$auto,  AB system) were corrected for foreground Galactic extinction using E(B -- V) obtained from Schlegel et al. (1998) and A$_{FUV}$ = E(B -- V) $\times$ 8.29, and A$_{NUV}$ = E(B -- V) $\times$ 8.18 (Seibert et al. 2005). We used the Cortese et al. (2008) method for computing the internal extinction for each object in the FUV band. For each A$_{FUV}$, we used Seibert et al. (2005) extinction law, shown above, to obtain the E(B-V). {\footnote{The values we find for E(B-V) obtained from spectroscopy for regions 3 and 4 of HCG~100 are slightly different from the values calculated from the UV images. This difference could be due to the fact that the slit is not sampling the entire region and might be missing some of the regions where the UV flux originates. E(B-V) values are: Region 3 (0.10+-0.01, 0.13) and Region 4 (0.09+-0.02, 0.05) - first value is from imaging, second value is from spectroscopy.}}

FUV and NUV colors were estimated using the task PHOT in IRAF, inside a fixed aperture of 4$\arcsec$ or 5$\arcsec$ radius, depending on the sizes of the sources, centered on the centroid of the light distribution of each NUV band detection. 

\subsection{Ages, masses and metallicities}

We used the method described in de Mello et al. (2008a, 2008b) to derive ages and masses from FUV and NUV GALEX images (Table \ref{table1}). 
For each region, ages were estimated using the models given by Starburst99 (SB99, Leitherer et al. 1999). These models were generated for an instantaneous burst, solar metallicity, Chabrier (2003) IMF, and are optimized for GALEX filters transmission curves. We have also generated models using Salpeter IMF and compared the results with Chabrier IMF. The difference in age between the two IMFs is around $\sim$ 2-4 Myr. In this paper we are presenting only the results generated with Chabrier IMF. We have also included in Table  \ref{table1} the errors for each age calculated from the errors in the colors. 

We compared the ages determined spectroscopically by Mendes de Oliveira et al. (2004) with our values using GALEX and found an excellent agreement for 3 of the 4 regions (\#1,\# 2 and \#4) for HCG~92. The object for which our age determination disagrees with respect to Mendes de Oliveira et al. (their region C and our region 7) is too close to another object and GALEX is not able to resolve them. We have also estimated ages from the  H$\alpha$ equivalent widths and SB99 models (Table \ref{table1}) and found agreement between the methods for 4 sources in HCG~92 and the 2 sources in HCG~100. However, 3 sources in HCG~92 are much older when calculated using the UV data than when using H$\alpha$. This could be due to the fact that ultraviolet light probes older stellar population than H$\alpha$ and/or due to other factors including wrong correction for dust attenuation and positioning of the slit. We will discuss this later in this session when we compare our estimates with other authors. We cannot exclude the possibility that density bounded conditions could produce lower equivalent widths,
that leading to apparent older ages when derived from them and SB99.

\begin{figure*}
\centering
\includegraphics[scale=0.6]{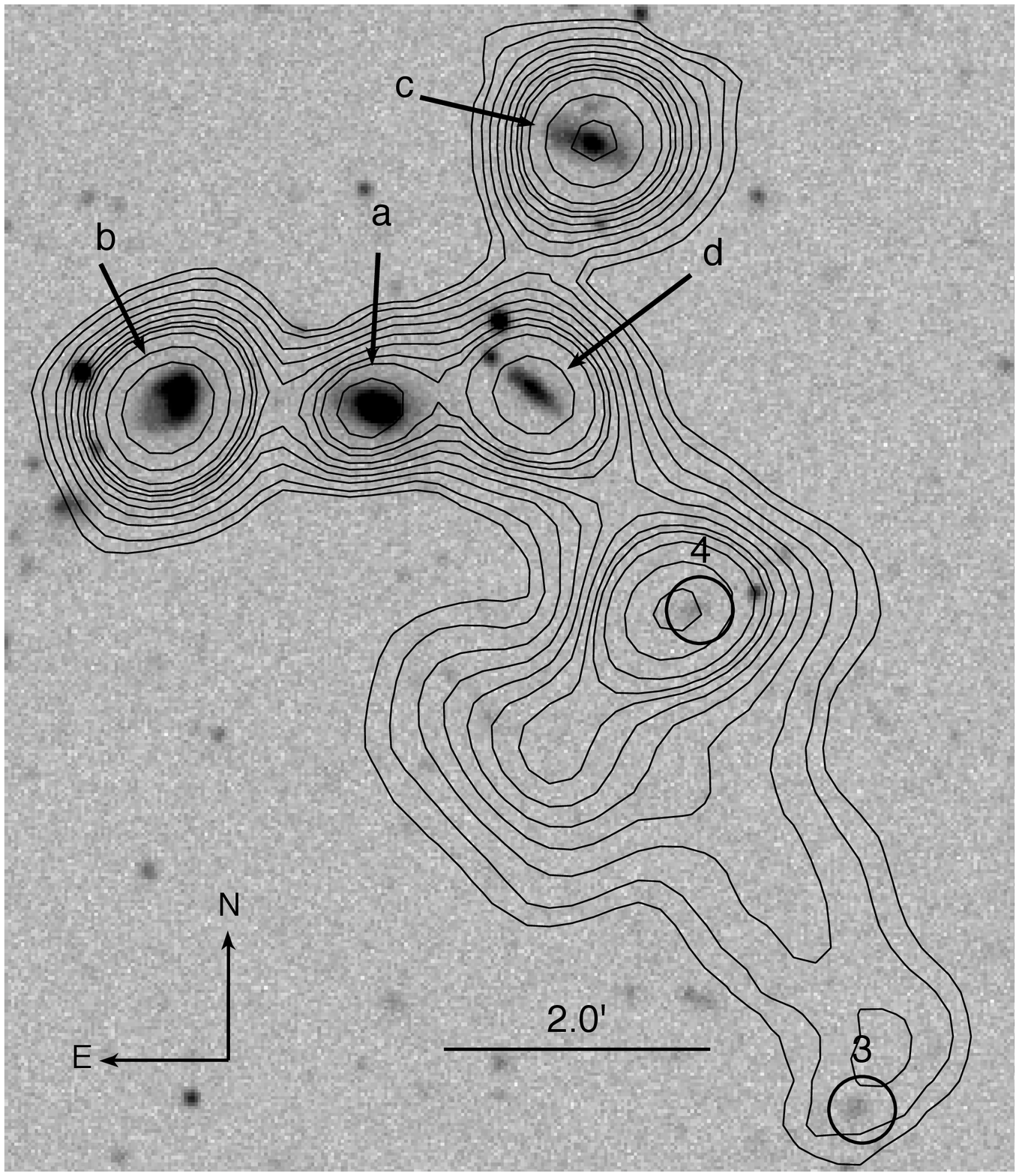}
\caption{NUV image of HCG~100 where galaxy members a, b, c and d are labeled. Two TDG candidates, \#3 and \#4, are marked. They fall within the H{\sc i} tail as shown in de Mello et al. (2008a). North is up and East is to the left. Bar length is 2$'$. VLA NHI contours are 0.6, 1.2, 2.1, 3.6, 4.4, 5.1, 5.9, 6.6, 7.4$\times$10$^{20}$ cm$^{-2}$. }\label{fig2}
\end{figure*}

\begin{table*}
\centering
\begin{minipage}[t]{\textwidth}
\scriptsize
\caption{Star-forming regions in the HI tail of HCG~92 and HCG~100}
\begin{tabular}{ccrccccccrc}
\hline
ID & ID$\_$region & R.A. & DEC. & V  & M$_B$\footnote{Calculated using magnitudes from Mendes de Oliveira et al. (2004) and de Mello et al. (2008a).} & 12+log(O/H)  & 12+log(O/H) & Log(M$_\ast$)\footnote{Stellar mass (M$_{\sun}$) obtained from Starburst99 monochromatic luminosity, L$_{\rm 1530}$ (erg/s/\AA), for the ages given in column 9. Stellar mass (M$_{\sun}$) for H100-\#3 and H100-\#4 are from de Mello et al. (2008).} & Age\footnote{Age (Myr) estimated from FUV-NUV color. Values given in brackets: ages estimated from H$\alpha$ equivalent width and SB99 models.} &  L$_{FUV}$ \\
 &  & 2000 & 2000 & (km s$^{-1}$) & mag & O3N2 & N2 & M$_{\odot}$ & Myr & erg s$^{-1}$\\
\hline
HCG~92 & 1\footnote{The respective ID's in Mendes de Oliveira et al. (2004) for regions 1,2, 4, and 7 in this table are d, a, b, and c.} & 339.04583  &33.98667  &  6615 $\pm$ 16 & -12.30 & 8.44$\pm$0.14 & 8.48$\pm$0.18 & 4.2  & 2.1$\pm^{1}_{3}$[5.4]     &  40.74 \\
      & 2$^d$ & 339.06250  & 33.98417 &  6613 $\pm$ 59 & -11.90 & 8.35$\pm$0.14 & 8.45$\pm$0.18  & 4.3  & 2.2$\pm^{2}_{5}$[4.1]      &  40.71\\
      & 3     & 339.05833 & 33.96750 &  6577 $\pm$ 69 & -11.58 & 8.45$\pm$0.14 & 8.55$\pm$0.18  & 4.1  & 2.4$\pm^{7}_{2}$[5.5]      &  40.52\\
      & 4$^d$ &  339.05833 & 33.97306 &  6659 $\pm$ 55 & -12.10 & 8.48$\pm$0.14 & 8.57$\pm$0.18  & 5.7  & 12.9$\pm^{3}_{5}$[5.5]     &  40.50\\
      & 5     & 339.99583 & 33.99417 &  6035 $\pm$ 19 & -11.34 & 8.40$\pm$0.14 & 8.44$\pm$0.18  & 7.6  & 50.1$\pm^{1}_{3}$[5.4]     &  41.30\\
      & 6     & 339.04583  & 33.95833 &  6543 $\pm$ 36 & -12.57 & ... & 8.69$\pm$0.18  & 8.5 & 100$\pm^{5}_{3}$[6.4]   &  41.67\\
      & 7$^d$     & 339.04583  &  33.98361 &  6612 $\pm$ 59 & -12.69 & 8.32$\pm$0.14 & 8.38$\pm$0.18  & 7.9 & 100$\pm^{2}_{4}$[2.44]      &  41.06\\
      & 8     & 339.07500  & 33.99333 &  5628 $\pm$ 17 & -13.23 & 8.50$\pm$0.14 & 8.60$\pm$0.18  & ... &...[7.6]  &  ...\\
      & 9     & 338.97500  & 33.95389 &  5780 $\pm$ 12 & -14.10 & 8.36$\pm$0.14 & 8.48$\pm$0.18  & ... &...[3.4]  &  ...\\
      & 10    & 338.98750  & 33.97972 &  6020 $\pm$ 81 & -15.25 & 8.62$\pm$0.14 & 8.67$\pm$0.18  & ... &...[5.3]  &  ...\\
      & 11    & 339.04583  & 33.98278 &  6553 $\pm$ 52 & -12.07 & 8.36$\pm$0.14 & 8.47$\pm$0.18  & ... &...[5.6]  &  ...\\
      & 12    & 339.04583  & 33.97389 &  6621 $\pm$ 71 & -9.88  & 8.46$\pm$0.14 & 8.55$\pm$0.18  & ... &...[3.9]  &  ...\\
HCG~100 & 3\footnote{ID from de Mello et al. (2008a).} & 0.27083 & 13.02333 & 5440 $\pm$ 61  & -14.54 & 8.43$\pm$0.14 & 8.43$\pm$0.18  & 4.7  & 1.0$\pm^{-}_{-}$[6.1]   &  40.46\\
       & 4$^{e}$ & 0.29167 & 13.08583 & 5337 $\pm$ 27  & -13.42 & 8.42$\pm$0.14 & 8.55$\pm$0.18  & 4.7  & 3.3$\pm^{3}_{1}$[2.1]    &  40.51\\
\hline
\vspace{-0.8cm}
\label{table1}
\end{tabular}
\end{minipage}
\end{table*}

\begin{figure}
\centering
\includegraphics[scale=0.32]{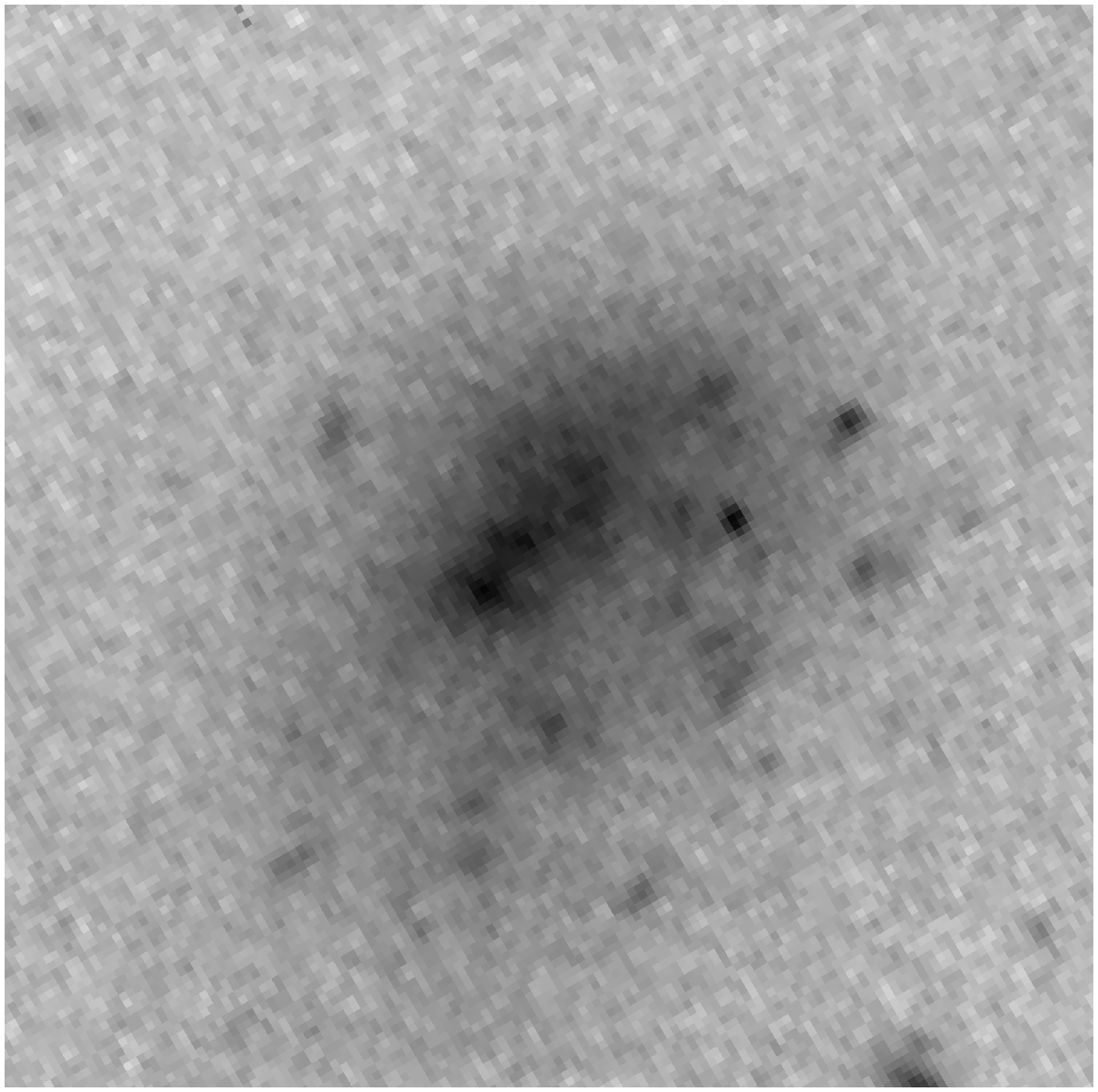}
\includegraphics[scale=0.32]{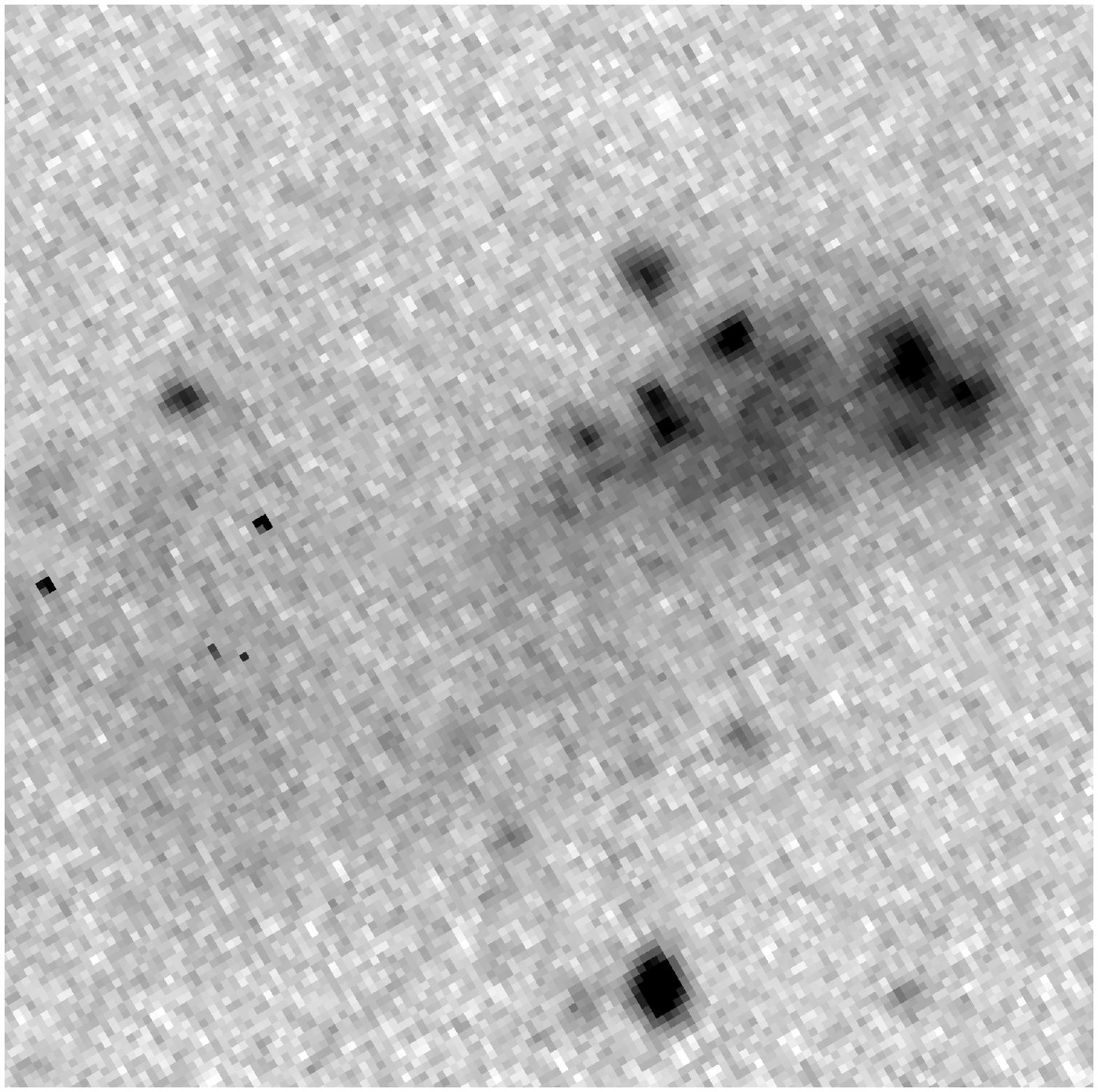}
\caption{Gemini image (filter r) of the two TDG candidates in the H{\sc i} tail of HCG~100 as originally identified in de Mello et al. (2008a) as objects \#3 (top) and \#4 (bottom). North is up and East is to the left.}\label{fig3}
\end{figure}

UV images showing the location of the newly detected members of HCG~92 and of HCG~100 are shown in Fig. \ref{fig1} and \ref{fig2}. 
The Gemini optical image (r filter; Fig.\ref{fig3}) shows the peculiar and knotty morphology of the two UV sources in HCG~100 which resemble dwarf galaxies. They are located in the H{\sc i} tail and we suggest that they might be TDG candidates. 

\begin{flushleft}
\begin{figure}
\hspace{-1.cm}
\includegraphics[scale=0.5]{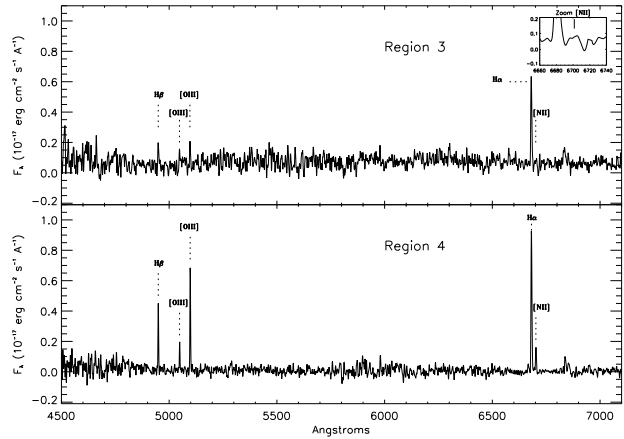}
\caption{Spectra of HCG~100's regions 3 (top) and 4 (bottom) identified in de Mello et al. (2008a). These spectra were taken with B600 and R400 gratings. The marked lines were used to estimate the oxygen abundance (12+log(O/H)). A zoom into the H$\alpha$ line region is shown on the right side of the top figure.}
\label{fig4}
\end{figure}\end{flushleft}

\begin{figure}
\hspace{-1.cm}
\includegraphics[scale=0.5]{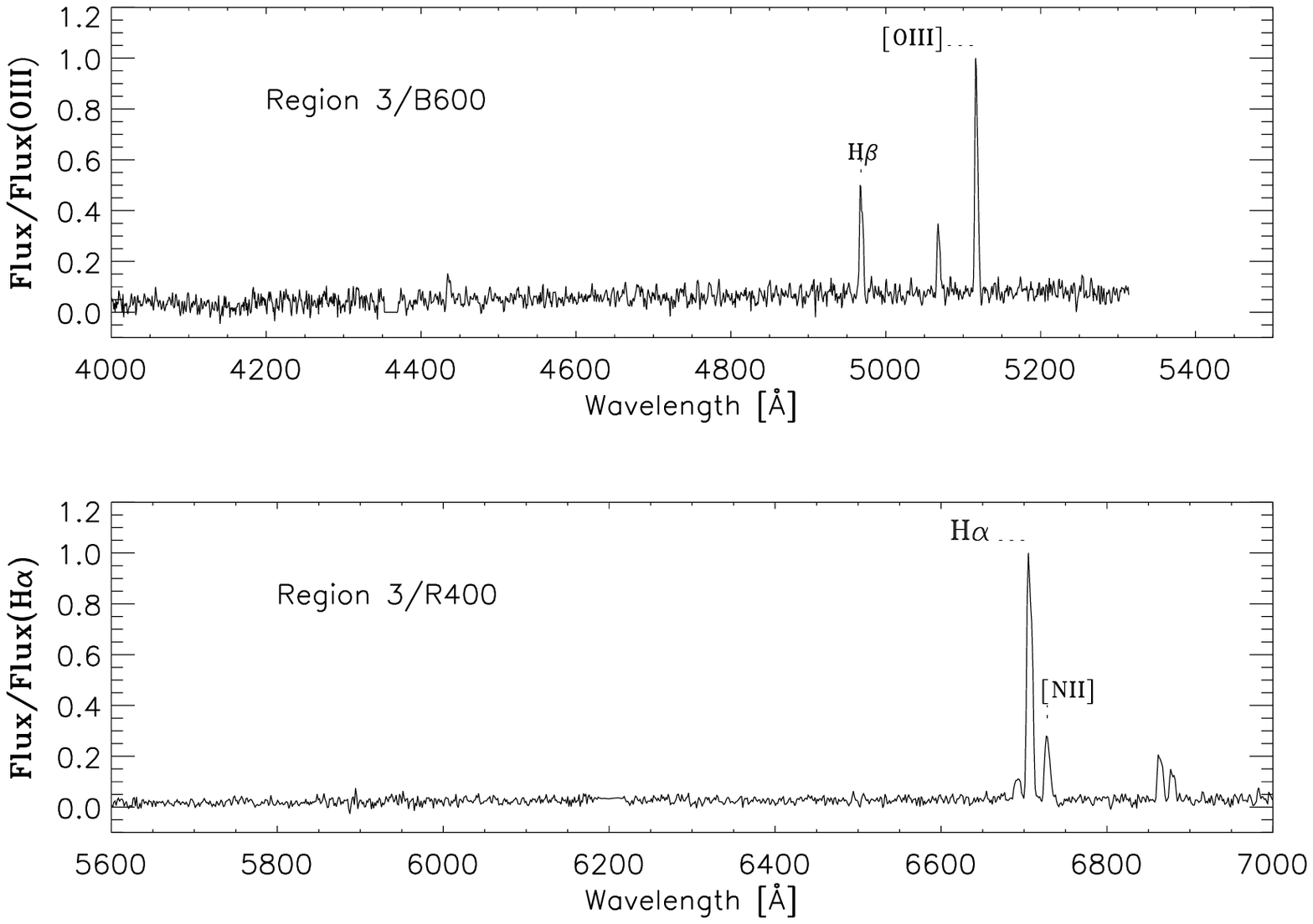}
\caption{Spectra of HCG~92's region 3. Spectra taken with B600 grating is in the top panel and R400 is in the bottom one. The marked lines were used to estimate the metallicities (12+log(O/H)). These spectra were not flux calibrated, given that no calibration star was available.}\label{fig5}
\end{figure}

\begin{figure}
\hspace{-1.cm}
\includegraphics[scale=0.5]{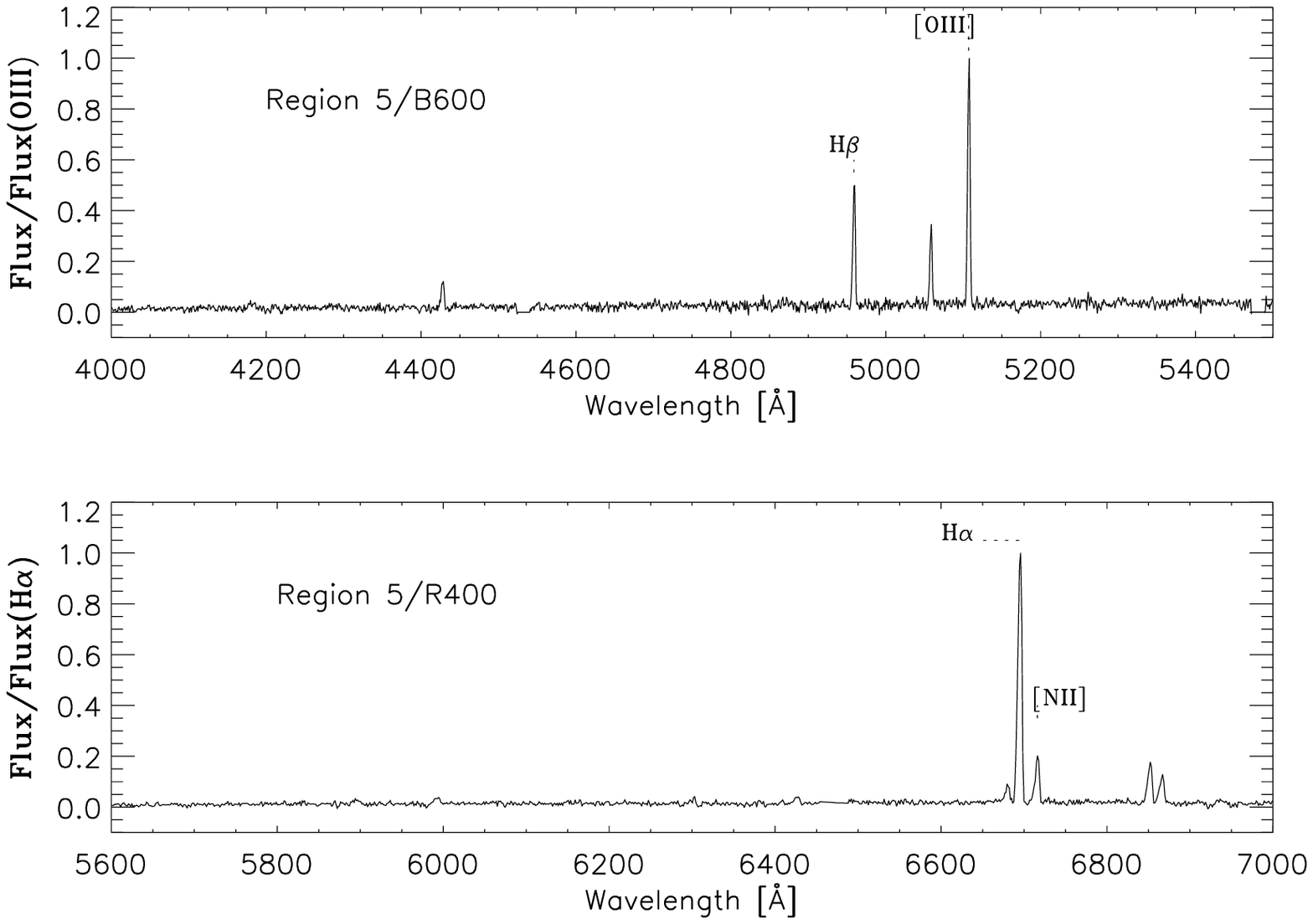}
\caption{Spectra of HCG~92's region 5. Spectra taken with B600 grating is in the top panel and R400 are in the bottom one. The marked lines were used to estimate the metallicities (12+log(O/H)). These spectra were not flux calibrated, given that no calibration star was available.}\label{fig6}
\end{figure}

In Fig. \ref{fig4} we show spectra taken with gratings R400 and B600 for the two TDG candidates of HCG~100, and in Fig.\ref{fig5}-\ref{fig6} we present two new star-forming regions of HCG~92, regions 3 and 5, which are  similar to the four objects described in Mendes de Oliveira et al. (2004).

Metallicities of the regions were calculated using the empirical methods O3N2 and N2, proposed and calibrated by Pettini and Pagel (2004).
These methods use line ratios 
 [O{\sc iii}] $\lambda$5007/H{\sc $\beta$} plus [N{\sc ii}] $\lambda$6584/H{\sc $\alpha$} and [N{\sc ii}] $\lambda$6584/H{\sc $\alpha$}, 
respectively, to estimate oxygen abundances. These estimators are adequate for faint extragalactic sources, such as the ones we are dealing with, because they are based only on very bright lines. The uncertainties on the calibration of these methods are 0.14 dex for O3N2 and 0.18 for N2 when 68\% of the points are included. Table \ref{table1} shows the estimated metallicities  for the regions in HCG~92 and HCG~100, plus our new results on velocities (measured from the emission lines), masses, and ages. As it can be noted in Table \ref{table1}, the metallicities derived from the O3N2 and the N2 methods are in close agreement. For one of the regions of HCG~92 (region 6), we used only the N2 method due to the lack of H$\beta$.

\begin{figure}
 \begin{center}
 \hspace{-1.cm}
 \includegraphics[scale=0.41]{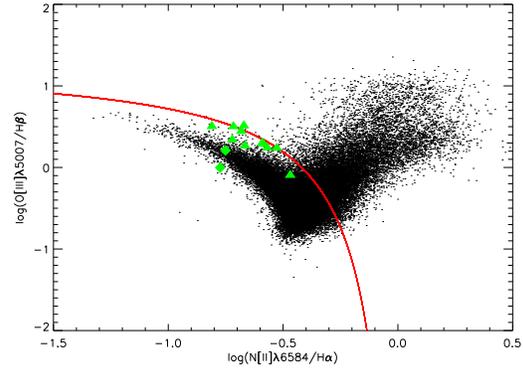}
  \caption{Line ratio diagram showing Sloan data from Kauffmann et al. (2003) and from star-forming regions in the HI tails of HCG~92 (triangles) and HCG~100 (diamonds). Objects with line ratios below the red line are classified as HII regions.  \label{bpt}}
\end{center}
  \end{figure}

We calculated line ratios and verified that all sources are star forming regions as shown in the BPT diagnostic diagram (Baldwin, Phillip \&Terlevich 1981) in Fig.\ref{bpt}. We plotted all Sloan sources from Kauffmann et al. (2003) for comparison.

\begin{figure}
 \begin{center}
 \hspace{-1.cm}
 \includegraphics[scale=0.41,angle=90]{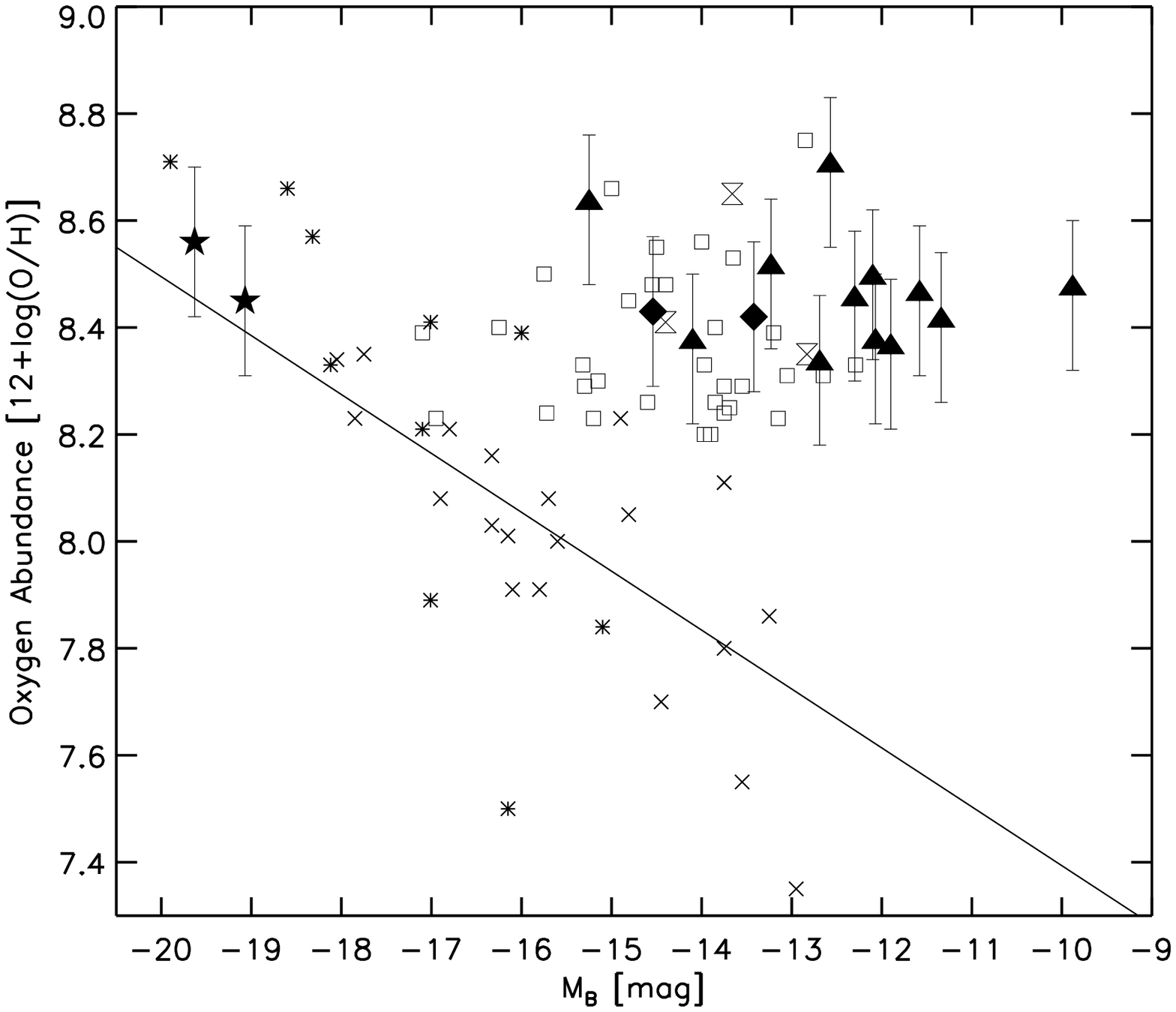}
  \caption{Absolute Magnitude (M$_{B}$) versus Oxygen Abundance for local isolated dwarf
 galaxies (crosses), knots in tidal features (open squares), main group members (asterisk), main members HCG~100 (stars), H{\sc ii} region of Croxall et al. (2009, Hourglass),  and our TDG candidates from HCG~92 (filled triangles) and HCG~100 (filled diamond). The line shows the correlation for dwarf galaxies from Weilbacher et al. (2003). The metallicities were obtained with O3N2 index for the majority of the regions. N2 calibrator was obtained for only one region of HCG~92. Adapted from Weilbacher et al. (2003). \label{fig7}}
\end{center}
  \end{figure}

In Fig.\ref{fig7} we show our new data plotted on the metallicity-luminosity diagram (adapted from Fig. 3 of Weilbacher et al. 2003). We estimated oxygen abundances using the method O3N2 described above. 
We also added new data from Croxall et al. (2009) to this figure.
As expected, HCG~100 main members occupy the left side of the diagram where luminous and metal-rich objects are located. Nearby dwarf irregular galaxies (Richer \& McCall 1995) follow the well-known correlation indicated by the linear fit.  Based on Fig.\ref{fig7}, all star-forming regions in the H{\sc i} tail of HCG~92 and HCG~100 have metallicities similar to those of knots in tidal features, i.e. they have metallicities higher than those of local dwarf galaxies. Lisenfeld et al. (2008) and Mendes de Oliveira et al. (2004) found similar results for star-forming regions in the intergalactic medium of the systems Arp~94 and HCG~92, respectively. 

Thus we conclude that the metallicity measurements obtained here allowed us to distinguish between ``classical'' dwarf galaxies and objects from tidal origin. We conclude that objects found within the H{\sc i} tails of HCG~92 and HCG~100 were formed by pre-enriched material, and their metallicities are similar or higher to that of their progenitor galaxies. It is also possible that their higher metallicities are due to a phenomenon called Infant Mortality (Fall et al. 2005) which destroys clusters by internal processes. 
In this way, the continuous star formation and destruction of stellar clusters could increase the metallicity of a given region after a few million years.

We have compared the coordinates of our targets with the ones identified in Trancho et al. (2012) using Hubble Space Telescope images and Gemini spectroscopy of HCG~92 and found no common sources. However, a close inspection of the HST images indicate that object T124 in Trancho et al. is the same as our object \#6, but had wrong coordinates quoted in their paper. The H{\sc i} tail is twice as long and has a different curvature than the optical tail, therefore HST small field of view did not cover the entire H{\sc i} tail. The work by Trancho et al. (see also Fedotov et al. 2012) focused only on the optical tail,  missing most of the targets we discovered. The age reported by those authors for T124 is in relative good agreement with the values we found for object \#6 using the Gemini spectra (1.5 Myr and 6 Myr, respectively). However, the age we found using the UV data, $\sim$100 Myr, is significantly higher.  This disagreement can be explained, as pointed out in Trancho et al., by the fact that the slit did not cover the entire complex and is missing the other components of the clump. The UV data on the other hand covers the entire region and is more representative of the cluster. The UV is also known for detecting older stellar population than H$\alpha$ and is a good age indicator for this type of stellar clusters.  Therefore, our results suggest that T124 (or \#6) is $\sim$100 Myr and 10$^{8.5}$ M$_{\odot}$, making it the most massive TDG candidate in the outskirts of HCG~92. We have also inspected the location of this object with respect to the H{\sc i} map (Mendes de Oliveira et al. 2004) and verified that it is located within one of density peaks which supports the idea that T124 (or \#6) is a TDG candidate. 
According to Trancho et al., two other stellar clusters close to T124, T117 and T122, are 7 and 50 Myr but are not resolved in the GALEX image and therefore are not part of our analysis. Another object, T118, is 125 Myr and has not passed our selection criteria (more details are given in the discussion Section 3). These 4 stellar clusters are within the optical tail which is estimated to have formed due to a close interaction between NGC 7318A and NGC 7319 $\sim$ 200 Myr ago (Renaud et al. 2010). Therefore, our data show that H{\sc i} tails in these two compact groups, HCG~92 and HCG~100, are laboratories of star formation.

\section{Discussion}


In order to explore whether the environment where the newly discovered stellar clusters and TDG candidates plays a significant role in their formation, we have analyzed other interacting systems with extended H{\sc i}  tails using a sample of galaxies from the Rogues gallery of H{\sc i} maps of peculiar and interacting 
galaxies compiled by J. Hibbard et al. (2001)\footnote[1]{http://www.nrao.edu/astrores/HIrogues/RoguesLiving.shtml}. 
We identified 25 interacting systems with GALEX data with exposure times $>$ 1 ksec which have also been observed by the Sloan Digital Sky Survey (SDSS). We followed the same method to identify UV sources and to obtain their ages as we did for the compact groups. Since we are looking for young regions predominately composed by the luminosity of O, B, and A stars, we defined a conservative cut in age at 100 Myr. We note that all new objects reported in the previous session for HCG~92 and HCG~100 are within this age range. 
We have also adopted a cut in luminosity equivalent to 100 O 8V stars (100 $\times$ L$_{\rm FUV}$ = 100 $\times$ 2 $\times$ 10$^{37}$ erg s$^{-1}$) or 4.2 $\times$ 10$^{39}$ erg s$^{-1}$. This limit was set based on the luminosities of stellar clusters found in the H{\sc i} bridge of M81 and M82  (a.k.a. ``blue blobs'') from de Mello et al. (2008b). 

Six of the 25 interacting systems originally selected have 35 UV sources with ages $<$ 100 Myr and luminosity higher than 4.2 $\times$ 10$^{39}$ erg s$^{-1}$ inside the H{\sc i} contours. A description of each these six interacting systems is given in the Appendix. 
In Table \ref{table_ig} we list the regions, their colors, ages, and luminosities, assuming that they are at the distance of the parent galaxies. Figs. \ref{fig2623} to \ref{fig4656} show the different systems with their H{\sc i} contours and the marked UV sources. We cannot exclude the possibility that a few or various of the UV sources we selected might be unrelated to the interacting galaxies, i.e. might be chance alignments. However, this method, when applied to HCG~100 by de Mello et al. (2008a), selected two TDG candidates within the H{\sc i} tail and several stellar clusters in M81/M82 (blue blobs, de Mello et al. 2008b). As shown in the previous session, our Gemini MOS data reveal that these two TDG candidates are at the same redshift as the galaxy group. Therefore, these two multi-wavelength studies show that this approach is successful in identifying UV sources which are either star-forming regions or TDGs related to the interacting systems with stripped H{\sc i} gas. 

\begin{figure}
\centering
\hspace{-1.cm}
\includegraphics[scale=0.41,angle=90]{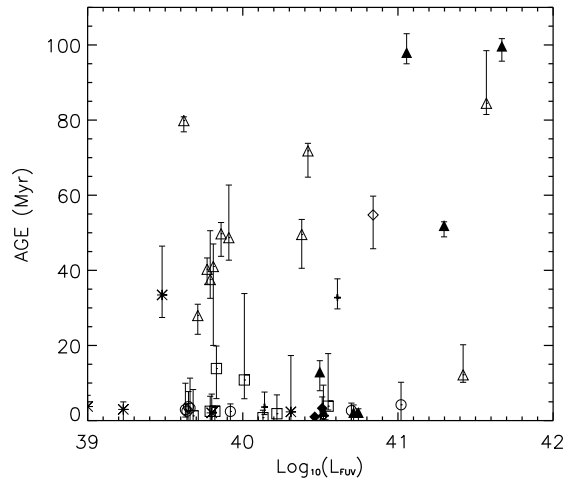}
\caption{FUV luminosity (erg s$^{-1}$) versus age (Myr) for star-forming regions outside galaxies in 6 interacting galaxies (different symbol),  
HCG~100 (filled diamond) and HCG~92 (filled triangles). \label{fig8}}
 \end{figure}

In order to compare the properties of the 35 UV sources found outside interacting galaxies with ones in HCG~92 and HCG~100 we have calculated their luminosities and searched for similarities in the two populations.  In Fig.\ref{fig8} we show the distribution of luminosities versus ages for the 35 regions and we verify that the population of intergalactic regions contains objects with luminosities as high as $\sim$ 10$^{42}$ erg s$^{-1}$ and as low as $\sim$ 10$^{39}$ erg s$^{-1}$, our lower limit. 
It is possible that the large range in luminosities indicates that we are dealing with different families of objects, as suggested in Mendes de Oliveira et al. (2004).  As seen in Fig.\ref{fig8} most of the regions in our sample are not as luminous as TDG candidates found in compact groups (filled symbols),  except for  object 2 around NGC~3079 and object 1 around NGC~3719. If that is the case, interacting galaxies are more likely to host star clusters while compact groups are more likely to host TDGs, as can be seen in Fig.\ref{fig8}. The low mass objects might be similar to the M81-M82 ``blue blobs''  described above. The high mass objects would then be the TDG candidates. The Kolmogorov-Smirnov probability test confirms that luminosities of stellar clusters in interacting systems are significantly different from the ones in compact groups (KS = 0.00013), i.e. luminosities are lower in the former than in the latter. We have also calculated masses using Starburst99 (Column 8 in Table\ref{table_ig}) and the same trend is verified.

Further spectroscopic observations of these regions, as we did for HCG~92 and HCG~100, are needed in order to confirm their membership and establish their metallicities.
It is possible that the group environment is more conducive to TDG formation (or better said, TDG survival) than pairs and mergers. This is in agreement with the simulations by Bournaud $\&$ Duc (2006) where specific conditions such as low impact velocity (v$<$ 250 km s$^{-1}$), prograde encounters and mass ratio up to 4:1 may lead to TDG formation. HCGs might harbor these conditions besides the possibility that group potential may be able to drive TDGs away from the nearby proximity of their progenitor galaxies. 

\section{Summary}

We presented new Gemini spectroscopy of fourteen star-forming regions within the H{\sc i} tails of HCG~92 and in HCG~100 confirming that they are at same redshifts of the groups. We estimated their metallicity and verified that they are metal-rich with respect to typical dwarf galaxies. This is possibly due to the fact that they were formed from pre-enriched material found in the intragroup medium. 

We analyzed GALEX FUV and NUV data of a comparison sample of 6 interacting galaxies containing a total of 35 UV sources in the H{\sc i} tails. These star-forming regions span a wide range of ages ($<$ 100Myr) and luminosity (10$^{39}$ --  10$^{42}$ erg s$^{-1}$). We compared their properties with those of the star-forming regions in the H{\sc i} tails of  HCG~92 and HCG~100. We concluded that they have on average lower luminosity than the ones in the H{\sc i} tails of compact groups. We suggest that this maybe is an environmental effect, i.e. that compact groups of galaxies with tidal tails of H{\sc i} are more likely to host more massive star-forming regions or TDGs than other interacting galaxies. Spectroscopy of these sources are needed in order to confirm that they are at the same redshift as the interacting galaxies and to establish their metallicities.

\subsection{Appendix}

We present the GALEX/FUV images of the 6 interacting systems with extended H{\sc i} gas. The FUV sources with ages $<$ 100 Myr and L$>$ 10$^{39}$ erg s$^{-1}$ are marked. The H{\sc i} contours are adapted from Hibbard et al. (2001).

\subsubsection{NGC~2623}

NGC~2623, also known as Arp~243, is located at 76.1 Mpc (Hattori et al. 2004). Bournaud et al. (2004) and Hattori et al. (2004) classified this object as a merger in an advanced stage. Hibbard \& Yun (1996) found that a large part of the HI gas is located far away from the stellar component of NGC~2623. 

\begin{figure}
\centering
\includegraphics[scale=0.25]{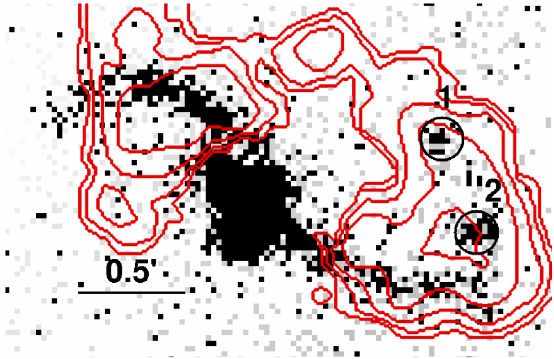}
\caption{FUV image of NGC~2623, regions with ages $<$ 100 Myr are marked with circles of 6'' radius. H{\sc i} contours=4 $\times$10$^{19}$cm$^{-2}$$\times$2$^{n}$ are from VLA C+D-array and provided by Hibbard et al. (2001).}\label{fig2623}
\end{figure}

In our analysis we detected two young UV sources in the western tail of NGC~2623, as show in Fig. \ref{fig2623}. One of them (region $\#$1 in Figure 5) seems to be associated with the giant H{\sc ii} region (which could be a TDG candidate) detected by Bournaud et al. (2004) in their Fabry-
Perot study. The age of this region is only 3 Myr. Its mass is about five times the mean mass of the intergalactic H{\sc ii} regions of Mendes de Oliveira et al. (2004). We detect another young region, $\#$2 (Figure \ref{fig2623}), within the HI contours of NGC~2623. Interestingly, this region is also detected in the H$\alpha$  map shown in Bournaud et al. (2004). The detection in the narrow band image confirms that these two regions belong to the NGC~2623 system. In Table \ref{table_ig} we list the main physical parameters for each object.

\begin{table*}
\centering
\begin{minipage}[t]{\textwidth}
\scriptsize
\caption{UV sources (ages $<$ 100 Myr) in H{\sc i}  tails of 6 interacting galaxies}
\begin{tabular}{ccccrccc}
\hline
ID & ID$\_${region} & R.A.J2000 & DEC.J2000 
&FUV-NUV\footnote{FUV and NUV magnitudes and errors were obtained using IRAF task phot (Poisson). The errors in the colors were calculated using the magnitude errors added in quadrature.}
&Age Myr\footnote{Age (Myr) estimated from FUV-NUV color.}
&L$_{FUV}$ (erg s$^{-1}$) &Log(M$_\ast$)\footnote{Stellar mass (M$_{\sun}$) obtained from Starburst99 monochromatic luminosity, L$_{\rm 1530}$ (erg/s/\AA), for the ages given in column 6.}\\
\hline
NGC2623 & 1  &  129.5834808  &   25.7584667  &  -0.14 $\pm$ 0.16 &    3.6$\pm^{4}_{2}$ &   40.14  & 4.7 \\
NGC2623 & 2  &  129.5805054  &   25.7511082  &   0.03 $\pm$ 0.11 &   32.7$\pm^{5}_{3}$  &   40.61 & 6.6 \\  
NGC3079 & 1   &   150.5596161  &   55.6378441   &   -0.43 $\pm$ 0.02  &       1.0$\pm^{1}_{4}$      &     40.52 & 4.3\\
NGC3079 &  2   &   150.5514526  &   55.5880432   &   -0.21 $\pm$ 0.12  &       3.0$\pm^{2}_{5}$    &     39.23 & 3.6\\
NGC3079 & 3   &   150.1344757  &   55.6083908   &   0.02 $\pm$ 0.04  &       33.4$\pm^{13}_{6}$   &     39.48 & 6.0\\
NGC3079 &  4   &   150.5026245  &   55.7128906   &   -0.19  $\pm$ 0.14 &        3.7$\pm^{3}_{8}$   &     39.95 & 3.8\\
NGC3079 &  5   &   150.4724426  &   55.7403946   &   -0.32 $\pm$ 0.13 &        2.1$\pm^{5}_{9}$   &     39.80 & 3.3\\
NGC3079 &  6   &   150.5094452  &   55.6856461   &   -0.22 $\pm$ 0.21 &       2.3$\pm^{15}_{5}$    &     40.31 & 3.8\\
NGC3359 & 1    &        161.5222015  &   63.1473923    &    -0.26 $\pm$ 0.21 &   2.7$\pm^{2}_{8}$  &     40.70   & 4.3\\
NGC3359 &  2    &        161.5304108  &   63.1609001    &    -0.13 $\pm$0.20 &   3.7$\pm^{4}_{1}$  &     39.65   & 3.5\\
NGC3359 &  3    &        161.5033569  &   63.2215042    &    -0.30 $\pm$ 0.18 &   2.4$\pm^{2}_{6}$  &     39.92   & 3.5\\
NGC3359 &  4    &        161.5233002  &   63.2346001    &    -0.09 $\pm$ 0.17 &   4.2$\pm^{6}_{8}$  &     41.02   &5.0\\
NGC3359 &  5    &        161.1769409  &   63.1001740    &    -0.23 $\pm$ 0.06  &    2.9$\pm^{7}_{2}$  &     39.63  &3.2 \\
NGC3359 &  6    &        161.7557220  &   63.2623863    &    -0.25 $\pm$ 0.2  &   2.5$\pm^{2}_{5}$  &     39.64  & 3.2 \\
NGC3359 &  7    &        161.6039429  &   63.2815933    &    -0.16 $\pm$ 0.13 &   3.3$\pm^{8}_{4}$  &     39.66  & 3.4 \\
NGC3627 & 1  &  170.1986084  &   12.9696226   &   0.67 $\pm$ 0.08   & 0.1$\pm^{5}_{9}$ &  41.43  & 5.2\\    
NGC3718 & 1    &        173.2548523   &   53.1872864    &      0.14 $\pm$ 0.19 &    84.5$\pm^{14}_{3}$   &   41.57 & 7.2 \\
NGC3718 &   2    &        173.2459717   &   53.1571503    &     0.05 $\pm$ 0.27 &    49.7$\pm^{3}_{6}$   &   39.86  & 5.8\\
NGC3718 &   3    &        173.2224731   &   53.1303406    &     -0.08 $\pm$ 0.06 &    12.2$\pm^{8}_{2}$   &   41.42 &5.9  \\
NGC3718 &   4    &        173.2071838   &   53.1464157    &     0.05 $\pm$ 0.13 &    49.5$\pm^{4}_{9}$   &   40.38 &5.7 \\
NGC3718 &   5    &        173.1886292   &   53.1595993    &     0.02 $\pm$ 0.12 &    41.0$\pm^{6}_{21}$   &   39.81 &5.0 \\
NGC3718 &   6    &        173.1446991   &   53.1508026    &     -0.02 $\pm$ 0.17 &    30.0$\pm^{3}_{5}$   &   39.71 &4.6 \\
NGC3718 &   7    &        173.0174866   &   53.1026154    &     0.11 $\pm$ 0.09 &    71.8$\pm^{2}_{7}$   &   40.42  &6.0\\
NGC3718 &   8    &        173.0266724   &   52.9917526    &     0.02  $\pm$ 0.18 &    40.3$\pm^{3}_{4}$   &   39.76  &4.9\\
NGC3718 &   9    &        173.1674805   &   52.9503860    &      0.13 $\pm$ 0.18 &    79.9$\pm^{1}_{3}$   &   39.62  &5.3\\
NGC3718 &   10   &        173.2238617   &   53.0397415    &     0.01 $\pm$ 0.14 &    37.5$\pm^{13}_{5}$   &   39.79 &4.9 \\
NGC3718 &   11   &        173.0760498   &   53.0345230    &     0.05 $\pm$ 0.12 &    48.7$\pm^{14}_{6}$   &   39.91 &5.2 \\
NGC4656 & 1  &      191.0442352  &   32.2628098   &     -0.05 $\pm$ 0.20 &   3.8$\pm^{14}_{1}$  &   40.55  & 4.4\\
NGC4656 &  2  &      191.0705109  &   32.2710037   &     -0.09 $\pm$ 0.20 &  10.8$\pm^{23}_{5}$  &   40.01  &5.1 \\
NGC4656 &  3  &      191.0773621  &   32.2755814   &     -0.10 $\pm$ 0.18 &   2.5$\pm^{4}_{12}$  &   39.79   &3.4\\
NGC4656 &  4  &      191.0822449  &   32.2876625   &     -0.24 $\pm$ 0.06 &   1.8$\pm^{5}_{15}$  &   40.22   &3.7\\
NGC4656 &  5  &      191.0815887  &   32.2807884   &     -0.26 $\pm$ 0.22 &   1.3$\pm^{7}_{2}$  &   39.68   &3.3\\
NGC4656 &  6  &      191.0652161  &   32.2919540   &     -0.07 $\pm$ 0.03 &  13.8$\pm^{6}_{8}$  &   39.83   &5.1\\
NGC4656 &  7  &      191.0808105  &   32.3068008   &     -0.31 $\pm$ 0.14 &   1.0$\pm^{2}_{4}$  &   40.13   &3.8\\
NGC4656 &  8  &      191.0811768  &   32.3120995   &     -0.09 $\pm$ 0.13 &   2.5$\pm^{1}_{6}$  &   39.82   &3.4\\
\hline
\vspace{-0.8cm}
\label{table_ig}
\end{tabular}
\end{minipage}
\end{table*}

\subsubsection{NGC~3079}

NGC 3079 is a giant spiral galaxy with two companions, MCG~9-17-9 (northeast) and NGC~3073 (southeast). NGC~3079 is located at 15 Mpc (de Vaucoulers et al. 1991) and it is a Seyfert 2/LINER with X-ray emission (Irwin \& Saikia 2003; Kondratko et al. 2005). This galaxy is one of the brightest observed mergers (Henkel et al. 1984). NGC~3073 is a dwarf galaxy with an elongated H{\sc i} tail aligned with the core of NGC~3079.

\begin{figure}
\centering
\includegraphics[scale=0.3]{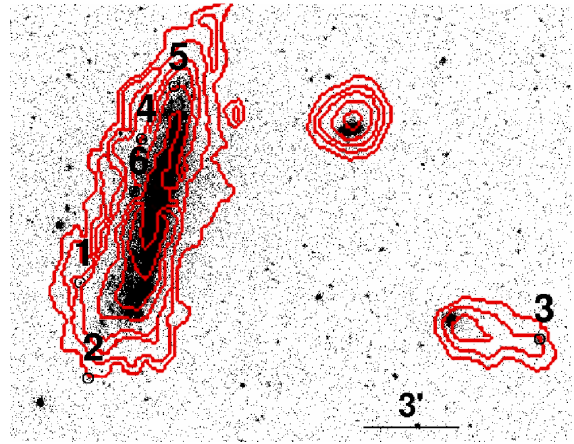}
\caption{FUV image of NGC~3079, regions with ages $<$ 100 Myr are marked with circles of 10'' radius. NGC~3073 is a weak UV source close to source \#3 in the figure.  H{\sc i} contours=(0.5, 1, 1.5, 2.4, 4, 10, 15, 25, 45) $\times$10$^{20}$cm$^{-2}$ are from VLA D-array and provided by Hibbard et al. (2001). }\label{fig3079}
\end{figure}

We found six UV sources (Fig. \ref{fig3079}) in this system. One of them seems to be associated with the H{\sc i} contours of NGC3073. In Table \ref{table_ig} we list the main physical parameters for each object. H{\sc i} contours

\subsubsection{NGC~3359}

NGC~3359 is a barred spiral galaxy with several spiral arms and an irregular morphology in the outer parts of the disk.

\begin{figure}
\centering
\includegraphics[scale=0.25]{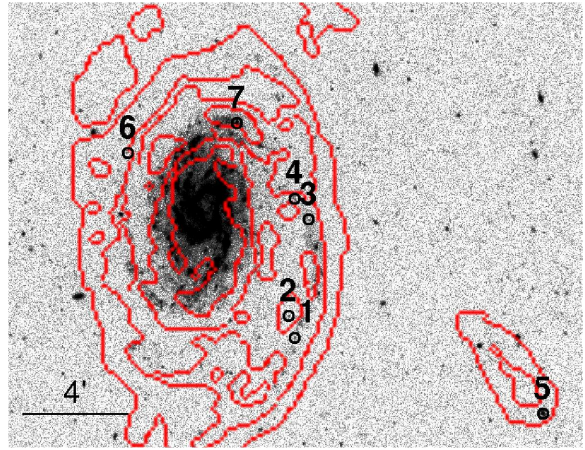}
\caption{FUV image of NGC~3359, regions with ages $<$ 100 Myr are marked with circles of 11'' radius. H{\sc i} contours=3 $\times$10$^{19}$cm$^{-2}$$\times$2$^{n}$ are from WSRT (30'' resolution). The most outer contour is 5$\times$10$^{18}$cm$^{-2}$ for 60'' resolution, and provided by Hibbard et al. (2001).}\label{fig3359}
\end{figure}

NGC~3359 is at 13.4~Mpc (Rozas 2008) and it shows strong arms in the UV which are not observed in the optical. We detected 7 UV sources (Fig. \ref{fig3359}) in this peculiar spiral, plus one UV source in the isolated HI cloud far from the disk (region 5).

\subsubsection{NGC~3627}

NGC~3627 is part of the Leo Triplet together with NGC~3623 and NGC~3628. The system is located at 6.7 Mpc (de Vaucouleurs 1975) and contain a remarkable H{\sc i} bridge and tail which can be due to an encounter between the galaxies in the past.

\begin{figure}
\centering
\includegraphics[scale=0.29]{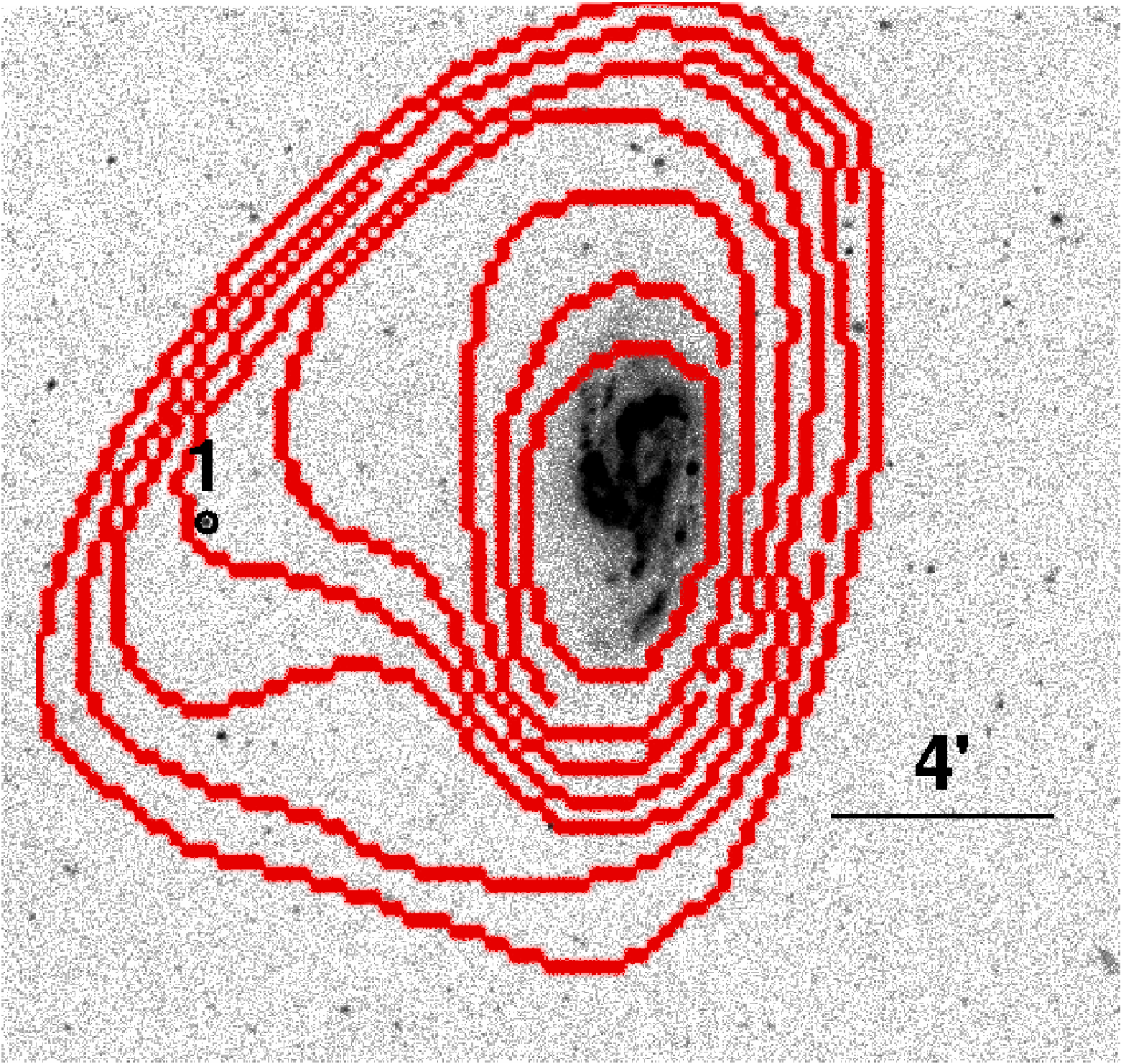}
\caption{FUV image of NGC~3627, regions with ages $<$ 100 Myr are marked with circles of 8'' radius. H{\sc i} contours= (3, 5, 10, 15, 25, 50, 75, 100, 200, 400, 600) K km s$^{-1}$ are from Arecibo and provided by Hibbard et al. (2001).}\label{fig3627}
\end{figure}

Here we present the results (Table \ref{table_ig}) only for NGC 3627 (Fig. \ref{fig3627}) since the other members are outside the GALEX field of view. We find only one UV source within the H{\sc i} contour, located close to where H{\sc i} seems to peak.

\subsubsection{NGC~3718}

NGC~3718 is part of the Great Bear group and it is at 17 Mpc (Tully 1998). This galaxy has a peculiar morphology, showing strong dust lanes and diffuse/peculiar spiral arms. It has a large extension of H{\sc i} gas, far from the optical disk (Allsopp 1979).

\begin{figure}
\centering
\includegraphics[scale=0.3]{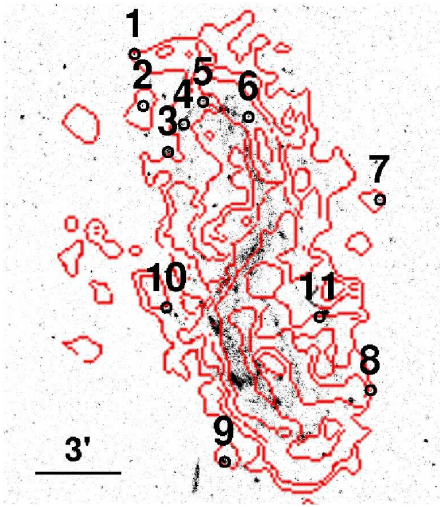}
\caption{FUV image of NGC~3718, region with age $<$ 100 Myr is marked with circle of 10'' radius. H{\sc i} contours =2 $\times$10$^{20}$cm$^{-2}$$\times$2$^{n}$ are from WSRT and provided by Hibbard et al. (2001).}\label{fig3718}
\end{figure}

NGC 3718 has 11 UV sources within the H{\sc i} contour in both arms of the galaxy and outside the R$_{25}$ optical radius (Fig. \ref{fig3718}, Table \ref{table_ig}).

\subsubsection{NGC~4656}

\begin{figure}
\centering
\includegraphics[scale=0.23]{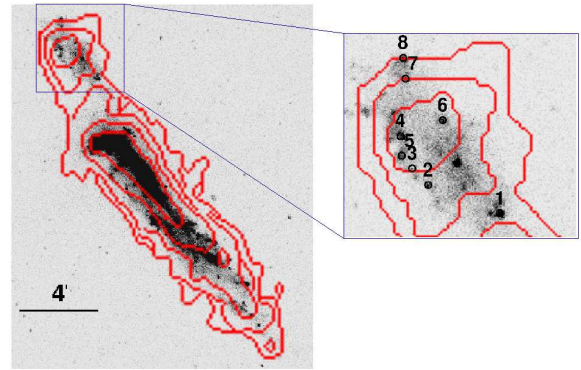}
\caption{FUV image of NGC~4656, and a zoom of the candidate to TDG, the regions with ages $<$ 100 Myr are marked with circles of 4'' radius. H{\sc i} contours =2 $\times$10$^{19}$cm$^{-2}$$\times$2$^{n}$ are from WSRT  and provided by Hibbard et al. (2001).}\label{fig4656}
\end{figure}

NGC 4656 is a spiral galaxy (Sc) interacting with NGC 4631 (Roberts 1968). They are linked by an H{\sc i}  bridge and are at 7.5 Mpc (Hummel et al. 1984). The bright region to the North of NGC 4656 resembles a TDG in the process of formation. We detected 8 UV sources in this area (Fig. \ref{fig4656} and Table \ref{table_ig}).

\section*{acknowledgements}

We are grateful to an anonymous referee for helpful comments and suggestions. 
DFdM was funded by NASA Research Grants NNG06GG45G and NNG06GG59G. F. U-V. acknowledges the financial support of FAPESP through an M. Sc. Fellowship, under contract 2007/06436-4 .C. M. d. O. acknowledges support from Brazilian agencies FAPESP (projeto tem\'atico 2006/56213-9), CNPq and CAPES. 
S. TÐF. acknowledges the financial support of FONDECYT (Chile) through a post-doctoral position, under contract 3110087 and FAPESP through the Doctoral position, under contract
2007/07973-3. This research has made use of the NASA/IPAC Extragalactic Database (NED) which is operated by the Jet Propulsion Laboratory, California Institute of Technology, under contract with the National Aeronautics and Space Administration. Some of the data presented in this paper were obtained from the Multimission Archive at the Space Telescope Science Institute (MAST). STScI is operated by the Association of Universities for Research in Astronomy, Inc., under NASA contract NAS5-26555. Support for MAST for non-HST data is provided by the NASA Office of Space Science via grant NAG5-7584 and by other grants and contracts.

\label{lastpage}
\end{document}